# Performance of random template banks

Bruce Allen[*]

*Max Planck Institute for Gravitational Physics (Albert Einstein Institute), Leibniz Universität Hannover,
Callinstrasse 38, D-30167, Hannover, Germany*

When searching for new gravitational-wave or electromagnetic sources, the $n$ signal parameters (masses, sky location, frequencies, etc.) are unknown. In practice, one hunts for signals at a discrete set of points in parameter space, called a template bank. These may be constructed systematically as a lattice or, alternatively, by placing templates at randomly selected points in parameter space. Here, we calculate the fraction of signals lost by an $n$-dimensional random template bank (compared to a very finely spaced bank). This fraction is compared to the corresponding loss fraction for the best possible lattice-based template banks containing the same number of grid points. For dimensions $n < 4$, the lattice-based template banks significantly outperform the random ones. However, remarkably, for dimensions $n > 8$, the difference is negligible. In high dimensions, random template banks outperform the best known lattices.

## I. INTRODUCTION

Many searches for gravitational-wave and electromagnetic signals are carried out using matched filtering, which compares instrumental data to waveform templates [1–3]. Because the parameters of the sources are not known *a priori*, many templates are required, forming a grid in parameter space [4–9]. Like the mesh on a fishing net, the grid needs to be spaced finely enough that signals do not slip through. But if the grid has far more points than are needed, the computational cost becomes excessive. For this reason, a substantial technology has evolved to create these grids [10–16]. What choice of template bank is best?

The traditional literature on the topic asserts that, for a fixed number of grid points, the optimal template bank is the one that minimizes the maximum distance (twice the covering radius) between any grid point and its closest neighbor [10,12,13,15,17–20]. However, as recently shown in Ref. [21], this is incorrect.

If the goal is to maximize the number of detections and the templates are closely spaced, then the optimal template bank minimizes the *average mismatch*: the average squared distance between any point in parameter space and the closest grid point. The bank which minimizes this quantity (at fixed grid point density) is called the optimal quantizer.

An extensive introduction to the topic of optimal quantizer lattices can be found in the remarkable book by Conway and Sloane [22], and an update on the current status can be found in Ref. [23].

Lattice-based template banks can be challenging to construct, particularly if the parameter-space metric is not flat. One solution is to build template banks by placing search grid points *at random* [13] in parameter space. Because they are simple and quick to construct, even in a curved parameter space, and because they can easily accommodate arbitrary parameter-space constraints and boundaries, such "random template banks" are appealing [14,24]. Note that random template banks may be improved by pruning away [25] grid points that are not needed. The result is then called a "stochastic template bank" [12].

Here, we provide a simple exact analysis of the performance of a random template bank. This analysis could have been done a decade ago, when such template banks were introduced [13]. However, the authors of Ref. [13] (following the mistaken conventional wisdom described above; see Ref. [21], Sec. IV) assessed the performance in terms of the covering radius [13] rather than in terms of the average mismatch.

Our analysis of random template bank performance has significant consequences. We find that in low dimensions, a random template bank performs poorly compared to a well-chosen lattice. However, as the dimension increases, the performance of a random template bank quickly approaches and then surpasses the performance of even the best lattices.

This paper assumes that the reader is familiar with Ref. [21] and is structured as follows. Section II defines the average mismatch $\langle r^2 \rangle$ in the usual quadratic approximation and reviews its relationship to the fraction of signals lost and to the scale-invariant second moment $G$ of a lattice. Section III defines a random template bank as a Poisson process in $n$ dimensions and calculates $\langle r^2 \rangle$ following an argument from Ref. [26]. This average mismatch is compared to that of the best currently known lattices and to that of the best theoretically possible lattices. In Sec. IV, we examine the assumptions implicit in the Poisson process and discuss the "dimensional reduction" case, where the template bank becomes "thin" in one or more dimensions. In Sec. V, we use results from Ref. [23] to calculate lost signals in template banks which are Cartesian products, since these are often used. In Sec. VI, we extend the results to cover the case of large mismatch, by replacing the normal quadratic approximation to the mismatch with the recently proposed spherical ansatz [27]. This is followed by a short conclusion.

The reader who is primarily interested in the results and not in the details should see Eq. (2.6) for the fraction $f$ of lost detections and then consult Fig. 1 and Table I. These show the performance of a random template bank, also comparing it to the best currently known lattice-based template banks and to the best theoretically achievable template banks.

## II. AVERAGE MISMATCH AND THE SECOND MOMENT $G$

As we have explained, the performance of a template bank is determined by the average mismatch [21]. For a given region of parameter space and a given number of grid points, this in turn is proportional to the scale invariant second moment $G$.

To define $G$ and show its relationship to the average mismatch, let $x \in \mathbb{R}^n$ be parameter-space coordinates, and let $\mathcal{V} \subset \mathbb{R}^n$ be the region of interest (for example corresponding to the desired ranges of masses and frequencies of interest in a search). Here, $x$ denotes a vector with $n$ Cartesian components, and we employ the standard Euclidean metric and norm.

[In general, the metric on parameter space has nonvanishing curvature. In such cases, the space can be subdivided into regions which are locally flat. In each region, new coordinates can be introduced, in which the transformed metric and norm take our assumed Euclidean form, as discussed in Ref. [21], following Eq. (2.6). A feature of random template banks is that no such construction is required for nonflat metrics [13].]

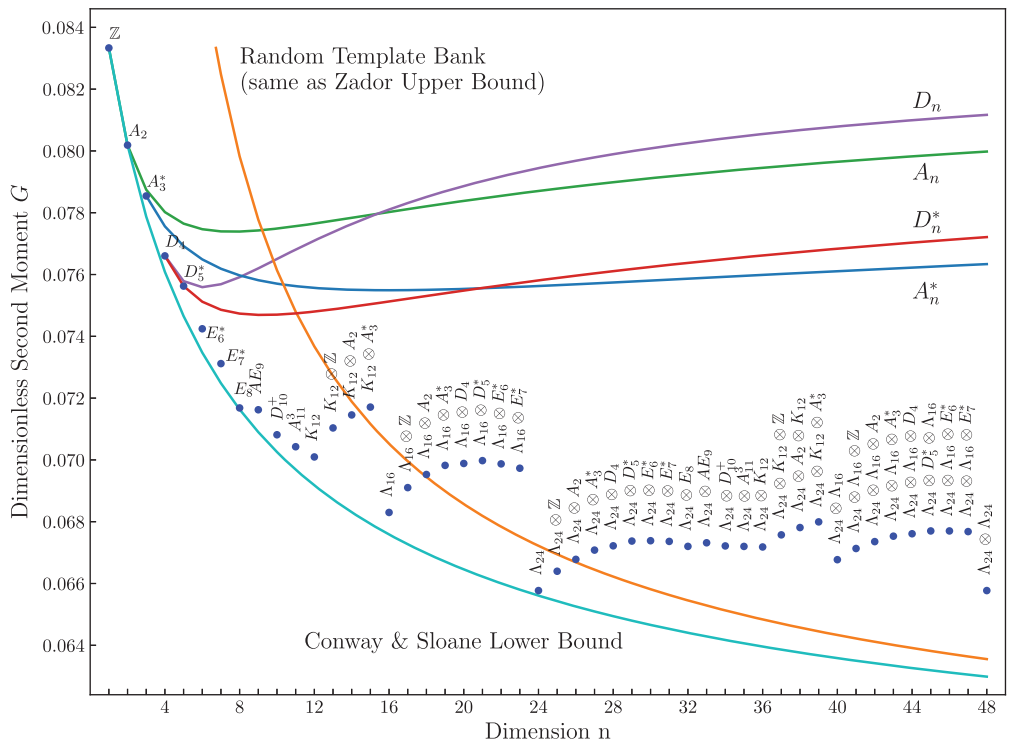

FIG. 1. The current record-holding (smallest $G$) lattice template banks (Ref. [23], Table 1) (blue points) lie above the conjectured Conway and Sloane [28] lower bound (cyan curve). The random template bank $G$ (orange) has its performance given by the Zador upper bound (3.8). The remaining colored curves show the well-known classical lattices $A_n$ and $D_n$ and their duals. For a fixed number of grid points, in dimensions $n > 8$, a random template bank has a performance (detection loss) which is within 10% of the theoretically best possible template bank (see Table I). In many higher dimensions (for example, 15 or 19), the random template bank outperforms *any* known lattice.

TABLE I. An ideal template bank has a loss factor $G$ at the Conway and Sloane lower bound $G_{CS}$, whereas a random template bank has a loss factor of $G_{\text{random}}$. The final column shows the fractional difference $(G_{\text{random}} - G_{CS})/G_{CS}$ in percent. For example, in $n = 9$ dimensions, if an ideal template bank were spaced to lose 5% of detectable signals, then a random template bank with the same number of grid points would lose about 5.5% of detectable signals (9.7% more).

| $n$ | $G_{CS}$ | $G_{\text{random}}$ | Max gain (%) |
|---|---|---|---|
| 1 | 0.08333 | 0.50000 | 500 |
| 2 | 0.08019 | 0.15915 | 98.5 |
| 3 | 0.07787 | 0.11580 | 48.7 |
| 4 | 0.07609 | 0.09974 | 31.1 |
| 5 | 0.07465 | 0.09132 | 22.3 |
| 6 | 0.07347 | 0.08608 | 17.2 |
| 7 | 0.07248 | 0.08248 | 13.8 |
| 8 | 0.07163 | 0.07982 | 11.4 |
| 9 | 0.07090 | 0.07778 | 9.7 |
| 10 | 0.07026 | 0.07614 | 8.4 |
| 11 | 0.06969 | 0.07480 | 7.3 |
| 12 | 0.06918 | 0.07367 | 6.5 |
| 13 | 0.06872 | 0.07272 | 5.8 |
| 14 | 0.06831 | 0.07189 | 5.2 |
| 15 | 0.06793 | 0.07116 | 4.8 |
| 16 | 0.06759 | 0.07053 | 4.3 |

The parameter-space $n$-volume is $V = V(\mathcal{V})$, where

$$V(\mathcal{S}) = \int_{\mathcal{S}} d^n x \tag{2.1}$$

is the volume of some subset $\mathcal{S} \subset \mathbb{R}^n$.

Suppose that $N$ search templates are located at grid points $x_1, \ldots, x_N$. Define the mismatch function

$$r^2(x) = \min(|x - x_1|^2, |x - x_2|^2, \ldots, |x - x_N|^2), \tag{2.2}$$

which is the squared distance from $x$ to the nearest template. For the given template bank, it is the fractional loss in (squared) signal-to-noise ratio (SNR) at each point in parameter space. The average of this quantity,

$$\langle r^2 \rangle = \frac{1}{V} \int_{\mathcal{V}} r^2(x) d^n x, \tag{2.3}$$

is the *average mismatch* [29].

Note that the fractional loss in (squared) SNR is only equal to $\langle r^2 \rangle$ when $\langle r^2 \rangle$ is small. The large $\langle r^2 \rangle$ case is treated in more detail in Sec. VI.

The goal of the template-bank architect is to minimize the average mismatch. This is because the fraction of signals which are lost (compared to a template bank with a very finely spaced grid) is given by Ref. [21], Eq. (5.6),

$$f = \frac{D}{2} \langle r^2 \rangle, \tag{2.4}$$

where $D$ is the effective dimension of the source distribution, which usually lies in the range $2 < D < 3$.

[The positive real constant $D$ describes the distribution of sources as a function of distance from the detector; see Ref. [21], Eq. (5.1). For example, Galactic pulsars with a planar distribution have $D = 2$; gravitational-wave binary inspiral sources distributed uniformly in volume have $D = 3$. Note that in this paper, to avoid confusion with the differential symbol, we use the symbol $D$ rather than the $d$ of Ref. [21].)

For example, suppose that 100 sources would in principle be detectable with a very finely spaced template bank and that these sources were distributed uniformly in space ($D = 3$). Then, a template bank with an $\langle r^2 \rangle = 3\%$ average mismatch loses about $f = 5\%$ of potential detections, so on average, 95 sources would be detected, and 5 would be lost.

To compare the relative performance of different template banks (i.e., different choices of the $N$ grid point locations $x_i$), it is convenient to define the *scale-invariant second moment*

$$G = \frac{1}{n} \frac{\langle r^2 \rangle}{(V/N)^{2/n}}. \tag{2.5}$$

Note that our definition in Eq. (2.5) is the conventional one [Ref. [22], Ch. 2, Eq. (87)], in spite of the appearance of $N$. This is because in the conventional definition, $V$ denotes the volume per grid point, which here is $V/N$.

The performance indicator $f$, which is the fraction of potentially detectable signals that are lost because of the discreteness of the template bank, may be expressed in terms of $G$, as

$$f = \frac{1}{2} n D (V/N)^{2/n} G. \tag{2.6}$$

Here, the "effective source dimension" $D$ is set by the spatial distribution of signal sources, and $V/N$ is the parameter-space volume per grid point.

To compare the performance of different template banks, fix the number of templates $N$, the parameter-space dimension $n$, and the volume of parameter space $V$. Then, the template grid with the smallest $G$ is the best choice, since it loses the smallest fraction of detections.

The simplest lattice, which is the $n$-dimensional cubic lattice, has a dimensionless second moment $G(\mathbb{Z}^n) = 1/12 \approx 0.08333$. A table showing the current records for the smallest $G$ among lattices (and also comparing the covering thickness) can be found in Ref. [21] and a larger and more recent table in Ref. [23]; these latter values are also shown in Fig. 1.

## III. RANDOM TEMPLATE BANKS

We now compute the performance of a random template bank. As first proposed by Ref. [13], a random template bank is created by randomly placing grid points with uniform probability within $\mathcal{V}$, locating each point independently of the positions of the other points. (Note that, while we continue to use the word "grid" to describe this template bank, the points are *not* uniformly spaced.)

In this construction, "uniform probability" means a Poisson process: the probability that an infinitesimal volume $dV$ contains a grid point is

$$P = \rho dV, \tag{3.1}$$

where the constant $\rho$ is the number of grid points per unit parameter-space volume $\rho = N/V$ or the number density of grid points. (This is easily generalized to a parameter space with nonflat metric; see the Conclusions and Ref. [13].)

By standard arguments for Poisson processes [Ref. [30], Eq. (14.24)], the probability of finding $\ell$ points within a finite volume $v$ is

$$P(\ell) = \frac{(\rho v)^\ell}{\ell!} e^{-\rho v}. \tag{3.2}$$

This assumption and its implications are examined more closely in Sec. IV.

We now calculate $\langle r^2 \rangle$ and $G$ following a beautiful argument [31] given by Torquato in Ref. [26]. Let $E(r)$ denote the *empty probability*. This is the probability that an $n$-ball of radius $r$, centered at a randomly selected point $x$ of parameter space, contains no grid points. The ball's $n$-volume is

$$V(B(r)) = \frac{\pi^{n/2}}{\Gamma(1+n/2)} r^n, \tag{3.3}$$

where the gamma function is defined by

$$\Gamma(z) = \int_0^\infty t^{z-1} e^{-t} dt \tag{3.4}$$

on the half-plane $\Re(z) > 0$ and by analytic continuation elsewhere. Setting $\ell = 0$ in Eq. (3.2), the empty probability is

$$E(r) = P(0) = e^{-\rho V(B(r))}. \tag{3.5}$$

Now, by definition, $E(r + dr)$ is the probability that a slightly larger ball of radius $r + dr$, randomly placed in parameter space, contains no grid points. This is a bit smaller than $E(r)$, and the difference,

$$E(r) - E(r + dr) = -\frac{dE}{dr} dr, \tag{3.6}$$

is the probability that the closest grid point to a random point $x$ lies in the shell of radius $r \in (r, r + dr)$ from $x$ [33].

Since $-\frac{dE}{dr} dr$ is the probability that the closest grid point lies in the shell of radius $(r, r + dr)$, it follows immediately that the average squared distance to the closest point in the template bank is

$$\begin{aligned}\langle r^2 \rangle &= -\int_0^\infty r^2 \frac{dE}{dr} dr \\ &= 2\int_0^\infty rE(r)dr \\ &= \frac{1}{\pi} \rho^{-\frac{2}{n}} \Gamma\left(1+\frac{n}{2}\right)^{\frac{2}{n}} \Gamma\left(1+\frac{2}{n}\right),\end{aligned} \tag{3.7}$$

where on the second line we have integrated by parts and on the third line we have substituted Eq. (3.5), changed variables, and used the definition Eq. (3.4) of the gamma function, along with $z\Gamma(z) = \Gamma(z+1)$.

The scale-invariant second moment $G$ of the random template bank follows from Eqs. (2.5) and (3.7), since $\rho = N/V$. This reproduces Ref. [26]'s Eq. (99) and furthermore, as noted by Torquato, gives exactly the Zador upper bound [34] for the optimal scale-invariant second moment

$$\begin{aligned}G_{\text{random}} &= G_{\text{Zador upper}} \\ &= \frac{1}{n\pi} \Gamma\left(1+\frac{n}{2}\right)^{\frac{2}{n}} \Gamma\left(1+\frac{2}{n}\right).\end{aligned} \tag{3.8}$$

This is plotted as the orange curve in Fig. 1.

As can be seen from Fig. 1 and Table I, the performance of a random template bank is very dependent upon dimension. In small dimensions, the performance is poor. For example, in one dimension, for a given parameter-space volume, signal source, and number of templates, a one-dimensional random template bank loses six times as many signals as the uniformly spaced grid $\mathbb{Z}$. In dimension 2, the random template bank loses almost twice as many signals as the hexagonal lattice $A_2$, and in dimension 3, it loses about 47% more signals than the optimal quantizer, which is the body-centered cubic (bcc, $A_3^*$) lattice. But the relative performance of a random template bank improves rapidly with dimension. By dimension 7, its performance is better than that of the hypercubic lattice $\mathbb{Z}^n$. By dimension 8, the random template bank loses only 11% more signals than the best known quantizer lattice $E_8$, which is likely optimal [35].

As the parameter-space dimension $n \to \infty$, both the Conway and Sloane conjectured lower bound and the Zador upper bound approach $G_\infty = 1/2\pi e \approx 0.058549$. In this sense, in higher dimensions, a random template bank, whose performance is equal to the Zador upper bound, is as good as one can get. In practice, this limit is quickly reached. If one selects a random template bank, then the final column of Table I shows the maximum fractional improvement (decrease from optimal) that is possible if there were a lattice that lies on the Conway and Sloane conjectured lower bound. This potential fractional improvement drops below 10% in dimension 8 and below 5% in dimension 15.

## IV. NUMBER OF TEMPLATES AND THE EFFECTIVE DIMENSION OF PARAMETER SPACE

Here, we examine in detail the assumptions made in the previous section and their implications regarding the shape and structure of the parameter space.

Consider Eq. (3.2) for a parameter space of finite volume $V$ containing exactly $N$ templates, and let $v = V$. Since this volume contains $N$ templates, we would expect to obtain $P(0) = 0$. But Eq. (3.2) instead gives $P(0) = e^{-\rho V} = e^{-N}$, which is nonzero for finite $N$.

Since our calculation only uses the probability $P(0)$, the results will hold in a parameter space with a finite volume $V$ if we assume that the number of templates $N$ is large. This is equivalent to requiring that $e^{-\rho V} = e^{-N} \ll 1$, which also ensures that truncating the Poisson distribution of Eq. (3.2) for $\ell > N$ has no significant effect.

To satisfy this condition, one may take the $n$-volume $V \to \infty$ with the density of grid points $\rho$ held constant. Alternatively, one may assume that the volume $v$ of the ball of radius $\sqrt{\langle r^2 \rangle}$ is small compared to $V$, so that $v/V \ll 1$.

A more subtle issue concerns the shape of the parameter space, as illustrated in Fig. 2. In Eq. (3.5), we have computed $P(0)$ from Eq. (3.2), which *assumes* that the ball $B(r)$ lies entirely within the volume $V$. But if we are near a parameter-space boundary, then that is no longer true. If the parameter space is "thick" in all dimensions, then such boundary effects can be neglected. On the other hand, if some regions of the $n$-dimensional parameter space are "thin", then their $n$-volume vanishes. The Poisson process then implies that they contain no template grid points. That would be correct if the probability that a source lies in those thin regions also has zero measure. However, in practice, such thin regions can occur in regions of parameter space where the detection statistic is (effectively) independent of one or more template parameters.

In such cases, the effective dimension of the parameter space is reduced. Although the $n$-volume vanishes, there is nevertheless a nonzero probability that a source lies in these thin regions, so a separate lower-dimensional template bank must be constructed to cover them. If this lower-dimensional template bank is a random bank, then it has a density $\rho'$ whose units (dimensions) differ from those of $\rho$. The value of $\rho'$ may be set via Eq. (3.7) to ensure that the average squared distance to the nearest point in the lower-dimensional parameter space is the same as that in the higher-dimensional space. For example, in the two-dimensional regions of Fig. 2, there are $\rho = 100$ grid points per unit area, implying that the average squared distance to the closest point is $\langle r^2 \rangle = 1/100\pi$. Equation (3.7) implies that a random grid along the one-dimensional line needs $\rho' = \sqrt{50\pi}$ grid points per unit length to obtain the same average spacing (mismatch).

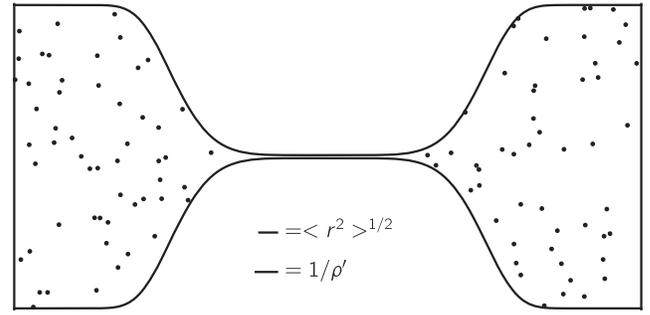

FIG. 2. A random template bank for a two-dimensional parameter space, with an (area) template point density $\rho = 100$. The upper bar has a length $\langle r^2 \rangle^{1/2} = (\pi \rho)^{-1/2}$, which is the root-mean-square (rms) distance to the closest template point. The middle of the parameter space has a region which is "thin" compared to this distance, with effective dimension 1. To obtain the same average mismatch $\langle r^2 \rangle$ as the two-dimensional part of parameter space, templates must be placed along this one-dimensional "line" with characteristic separation $1/\rho' = (\pi \rho/2)^{-1/2}$ corresponding to the lower bar.

The degree of dimensional reduction, i.e., the number of thin dimensions, is determined by the density of templates. A dimension is thin if its characteristic length scale is comparable to, or less than, the distance $\sqrt{\langle r^2 \rangle}$ given by Eq. (3.7). If the template density is made very high, then none of the dimensions is thin, but as the template density is lowered, $\sqrt{\langle r^2 \rangle}$ increases, and eventually the parameter space becomes thin in some dimensions.

## V. PRODUCT TEMPLATE BANKS

It is often desirable to form a template bank as the Cartesian product of two lower-dimensional template banks. For example, this occurs in searches where one of the parameter-space dimensions is frequency, and SNR values are obtained via a fast Fourier transform (FFT) from time-domain data. Such an FFT yields evenly spaced frequency bins. A second example is binary inspiral searches, where one of the parameter-space dimensions is binary coalescence time, sampled at the same sample rate as the data. In both examples, the parameter-space grid has a factor which is the evenly spaced one-dimensional lattice $\mathbb{Z}$.

In the most general approach to such cases, the template bank on the full $n$-dimensional parameter space is the Cartesian product of two lower-dimensional template bank "factors," whose dimensions are $n_a$ and $n_b$, with $n = n_a + n_b$. Recent work [23] shows how the relative grid spacings of the two factors can be scaled or adjusted to achieve the smallest possible value of $G$ for the resulting product. After that scaling, the product template bank has a scale-invariant second moment given by [see Ref. [23], Eq. (41)]

$$G = G_a^{\frac{n_a}{n}} G_b^{\frac{n_b}{n}}, \tag{5.1}$$

where $G_a$ and $G_b$ are the scale-invariant second moments of the two factors. Since $G(\mathbb{Z}) = 1/12$, the one-dimensional examples above correspond to $n_a = 1$ and $G_a = 1/12$.

In this way, the results of this paper can also be used to characterize and optimize the performance of template banks that are constructed as a product of a random template bank with a lattice or of two independent random template banks.

## VI. LOSS FRACTION AT LARGE MISMATCH

Up to this point in the paper, we have only considered "closely spaced" random template banks. We now generalize those results to arbitrarily large spacing. To distinguish these two cases, it is helpful to define

$$\Delta = \rho^{-1/n} = \left(\frac{V}{N}\right)^{1/n}, \tag{6.1}$$

which is the *characteristic distance* between grid points.

If the templates are closely spaced, then $\Delta$ is small. From Eqs. (2.4) and (3.7), this ensures that the fraction of lost signals

$$f = \frac{D}{2\pi}\Gamma\left(1+\frac{n}{2}\right)^{\frac{2}{n}}\Gamma\left(1+\frac{2}{n}\right)\Delta^2 \tag{6.2}$$

is small: $f \ll 1$. However, the treatment in Sec. III clearly breaks down if the grid spacing $\Delta$ becomes too large, since in that case the loss fraction $f$ in Eq. (6.2) would exceed unity. This is inconsistent, since by definition $f \leq 1$. This inconsistency arises because Sec. III assumes the "quadratic approximation" to the mismatch, which is invalid for large separations.

In this section, we make use of the "spherical ansatz" of Ref. [27] to compute the loss fraction of a random template bank for arbitrarily large template grid spacing $\Delta$. As before, the calculation for a random template bank is much simpler than for a lattice.

Employing the spherical ansatz, the loss fraction of Eq. (2.4) $f = D\langle r^2\rangle/2$ is replaced by

$$f = \langle s(r)\rangle, \tag{6.3}$$

where

$$s(r) = \begin{cases} 1 - \cos^D r & \text{for } r \leq \pi/2, \text{ and} \\ 1 & \text{for } r > \pi/2. \end{cases} \tag{6.4}$$

[These equations are derived in Ref. [27], Eq. (5.10), and Ref. [38], Eqs. (3.6) and (3.7). When $r$ is small, expansion of Eq. (6.4) in a Taylor series for small $r$ gives $s(r) \approx Dr^2/2$, recovering Eq. (2.4).]

To calculate $\langle s(r)\rangle$, we proceed as in Sec. III, beginning with Eq. (3.7), to obtain

$$\begin{aligned}\langle s(r)\rangle &= -\int_0^\infty s(r)\frac{dE}{dr}dr \\ &= \int_0^\infty \frac{ds(r)}{dr}E(r)dr \\ &= \int_0^{\pi/2} E(r)\frac{d}{dr}(1-\cos^D r)dr.\end{aligned} \tag{6.5}$$

In the second line, we have integrated by parts, since $s(r)$ vanishes at $r=0$ and $E(r)$ vanishes as $r \to \infty$. The third line follows because the derivative of $s(r)$ vanishes for $r > \pi/2$.

To compute this in closed form, we rewrite the integral in terms of the "expected values" of even powers of $r$. (These are defined as in Ref. [21], Eq. (5.11), with the caveat that the corresponding integrals are truncated at $r = \pi/2$. To emphasize this, we use $R$ rather than $r$ inside the angle brackets.) Thus, we define the *truncated moments*

$$\begin{aligned}\langle R^p\rangle &= \int_0^{\pi/2} E(r)\frac{d}{dr}r^p dr \\ &= \frac{p}{n}\Delta^p \pi^{-\frac{p}{2}}\Gamma\left(1+\frac{n}{2}\right)^{\frac{p}{n}}\gamma\left(\frac{p}{n},\frac{\pi^{\frac{3}{2}n}}{2^n\Delta^n\Gamma(1+\frac{n}{2})}\right),\end{aligned} \tag{6.6}$$

where the *lower incomplete gamma function* is defined by

$$\gamma(z,x) = \int_0^x t^{z-1}e^{-t}dt. \tag{6.7}$$

To use these moments to compute the loss fraction from Eq. (6.5), first expand $\cos^D r$ in a Taylor series, and then replace the (even) powers of $r$ using Eq. (6.6). One obtains the loss fraction $f = \langle s(r)\rangle$ given by

$$\begin{aligned}f = \frac{D}{2}\langle R^2\rangle &- \frac{3D^2-2D}{24}\langle R^4\rangle + \frac{15D^3-30D^2+16D}{720}\langle R^6\rangle \\ &- \frac{105D^4-420D^3+588D^2-272D}{40320}\langle R^8\rangle + \cdots.\end{aligned} \tag{6.8}$$

The loss fractions $f$ for random template banks are shown in Fig. 3 for a $D = 3$-dimensional source distribution. Note that, while Eq. (6.8) does not show the expansion terms proportional to $\langle R^{10}\rangle$ and $\langle R^{12}\rangle$, these are nevertheless included in Fig. 3, providing accuracy substantially greater than the plotting line width.

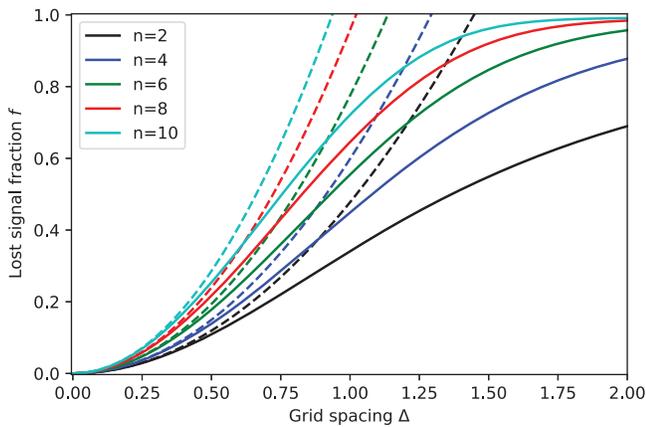

FIG. 3. The fraction of signals which are lost by a random template bank as a function of the grid spacing $\Delta$. These are computed using the spherical ansatz for the mismatch, for a $D = 3-$dimensional source distribution; the curves show parameter-space dimensions $n = 2, 4, \ldots, 10$. The dashed lines show the quadratic approximation for the mismatch Eq. (6.2), which is accurate at small grid spacings.

## VII. CONCLUSION

Random template banks are practical to employ because they are quick and simple to construct. It is remarkable that their performance is so easily characterized.

This analysis would have been possible when random template banks were first introduced in Ref. [13]. However, as we have explained, the authors of that work were focused on the covering radius or, more strictly speaking, on the "effective covering radius." Here, "effective" means that a specified (large) fraction of the parameter space was within a region covered by balls of the specified radius. This approach was necessary because the covering radius is defined by the first positive root of the empty probability $E(r)$. But, as can be seen from Eq. (3.5), in the case of a random template bank, $E(r)$ has no positive roots. Hence, the authors of Ref. [13] made use of an effective covering radius, at which $E(r)$ had decreased to an acceptably small value. This leads to a more complex treatment than the one given here.

For simplicity in this short paper, we have concentrated on the simplest case, with a flat parameter-space metric. However, these results also apply to the nonflat case, provided that the density of grid points is large enough to ensure that the signal manifold around each grid point is well approximated by flat space in the vicinity of the nearest neighboring $n$ grid points. If so, then a Poisson random template bank may be created by placing grid points with a constant probability density per unit volume $dV = \sqrt{\det(g_{ab})} d^n x$, which is the volume measure induced by the parameter-space metric [13]. This could also be modified to account for a varying probability of sources, as in Ref. [21] Sec. VI.

Random template banks outperform cubic lattices in dimensions $n > 7$ and are within 10% of optimal for dimension $n > 9$. However, it is currently not practical to carry out blind searches in such high dimensions because so many templates are needed. To date, the largest number of templates employed in (continuous) gravitational-wave searches (Ref. [39], Sec. IV.1) is of order $10^{18}$, in a four-dimensional parameter space. However, we expect that advances in quantum computing technology will eventually permit corresponding higher-dimensional searches [40]. For these, random template banks might be the best approach, or close enough to be equivalent.

For the moment, as we have explained, random template banks have been employed [41–44] for practical reasons. In such applications, the results of this paper are of interest because they permit the performance of such banks to be predicted in advance. In addition to being used for constructing and characterizing random template banks, these results may also be used to characterize any type of template bank, yielding a quantitative assessment of the degree to which that template bank has improved on a random bank. The results here may also be applied to characterize "injection studies" which are used to assess data analysis pipelines. In these, simulated signals picked from random locations in parameter space are added to detector output and passed through an analysis pipeline.

## ACKNOWLEDGMENTS

I acknowledge Erik Agrell, Daniel Pook-Kolb, and Andrey Shoom for many interesting discussions about lattices; Chris Messenger, Ben Owen, Maria Alessandra Papa, Reinhard Prix, and B. S. Sathyaprakash for many interesting discussions about template banks; and Salvatore Torquato for helpful correspondence. I also thank the anonymous referee, whose helpful comments about dimensional reduction led to the addition of Sec. IV, and Chris Messenger for detailed comments on the manuscript.